\documentclass[aps,prl,twocolumn,groupedaddress,showpacs]{revtex4}

\usepackage{graphicx}
\usepackage{dcolumn}
\usepackage{bm}

\begin{document}

\title{Magnetic Fluctuations of Filled Skutterudites \\
Emerging in the Transition Region between Singlet and Triplet States}

\author{Takashi Hotta}

\affiliation{Advanced Science Research Center,
Japan Atomic Energy Research Institute,
Tokai, Ibaraki 319-1195, Japan}

\date{\today}

\begin{abstract}
In order to clarify magnetic properties of filled skutterudites,
we analyze the Anderson model including seven $f$ orbitals hybridized
with an $a_{\rm u}$ conduction band using a numerical technique.
For $n$=2 corresponding to Pr-based filled skutterudites,
where $n$ is the local $f$-electron number,
even if the ground state is a singlet,
there remain significant magnetic fluctuations from
a triplet state with a small excitation energy.
This result can be understood by the fact that $f$-electron states
are clearly distinguished as itinerant and localized ones
in the filled skutterudite structure.
This picture also explains the complex results for $f$-electron
magnetic susceptibility and entropy for $n$=1$\sim$13.
\end{abstract}

\pacs{74.25.Ha, 75.20.Hr, 74.70.Tx}


\maketitle


The recent discovery of heavy fermion superconductivity
in PrOs$_4$Sb$_{12}$ \cite{PrOs4Sb12} has triggered
intensive studies on filled skutterudite compounds
in the research field of condensed matter physics.
Among the crystalline electric field (CEF) states of
filled skutterudites characterized by the
$T_{\rm h}$ point group \cite{Takegahara}, a
$\Gamma_1$ singlet has been found experimentally to
be the ground state of PrOs$_4$Sb$_{12}$ \cite{Gamma1}.
However, the results of NMR experiments \cite{Kotegawa}
suggest that the superconductivity is unconventional.
In fact, PrOs$_4$Sb$_{12}$ exhibits exotic features
such as multiple superconducting phases \cite{Izawa}
and the breaking of time-reversal symmetry as detected
by $\mu$SR experiments \cite{Aoki1}.
On the other hand, in the same family of Pr-based filled skutterudites,
PrRu$_4$Sb$_{12}$ is a conventional $s$-wave superconductor \cite{Yogi},
PrOs$_4$P$_{12}$ is a non-magnetic metal \cite{Sekine1},
PrRu$_4$P$_{12}$ exhibits a metal-insulator transition \cite{Sekine1},
and PrFe$_4$P$_{12}$ shows exotic quadrupolar ordering \cite{Aoki2}.
It is quite interesting that the electronic properties are easily changed
by simple substitution of transition metal atoms and pnictogens.

Besides Pr-based filled skutterudites,
many other kinds of filled skutterudite
compounds have been also synthesized.
Those materials also exhibit a variety of electronic properties:
La-based filled skutterudite materials are known to be conventional
BCS superconductors.
Ce-based filled skutterudites are Kondo semiconductors
with energy gaps up to a thousand Kelvins.
For skutterudites containing rare-earth ions other than
La, Ce, and Pr, a ferromagnetic ground state has been
frequently observed, as for instance in RFe$_4$P$_{12}$ with
R = Nd, Sm, Eu, Gd, Tb, Dy, and Ho.
However, antiferromagnetic ground states have also been found
in GdRu$_4$P$_{12}$ and TbRu$_4$P$_{12}$ \cite{Sekine2}.
It is clearly desirable to identify the key issues which control
the electronic properties of filled skutterudite compounds
from a microscopic point of view.

However, a microscopic theory for magnetism and superconductivity of
$f^n$-electron systems with $n$$>$1 ($n$ is local $f$-electron number)
has not been satisfactorily developed up to now,
owing to the complexity of the many-body problem which stems from
the competition among strong spin-orbit coupling, Coulomb interactions,
and the CEF effect.
To include such interactions, the $LS$ coupling scheme has been
widely used.  However, that approach cannot exploit standard quantum-field
theoretical techniques,
since Wick's theorem does $not$ hold in the $LS$ coupling scheme.
In order to overcome such a difficulty, we have proposed to
construct a microscopic model for $f$-electron systems
based on a $j$-$j$ coupling scheme \cite{Hotta1}.
When we attempt to apply the latter model
to filled skutterudites, there are several issues to be addressed.
In particular, one may have doubts about
the application of the $j$-$j$ coupling
scheme to rare-earth materials, since Coulomb interactions are
generally larger than the spin-orbit coupling
in $4f$-electron systems.
It is also necessary to clarify how the CEF energy levels
for $f^n$-electron systems are reproduced
in the $j$-$j$ coupling scheme.

In this Letter, we analyze a multi-orbital Anderson model,
which correctly contains spin-orbit coupling, Coulomb interactions,
and the CEF effect.
For $n$=2, significant magnetic fluctuations are found to remain
even in an $f^2$-electron system with a $\Gamma_1$ singlet ground state,
since there is a $\Gamma_4^{(2)}$ excited state triplet
with a small excitation energy
controlled by the CEF interaction.
It is also shown that essentially the same result is
obtained in an Anderson model constructed
with the $j$-$j$ coupling scheme.
We find that in filled skutterudites, a certain $f$ orbital
hybridizes with the conduction band,
while other orbitals are localized.
This picture also explains the magnetic properties of
filled skutterudites for $n$=1$\sim$13.


The Anderson Hamiltonian is written as
\begin{eqnarray}
  \label{AndersonModel}
  H = \sum_{\bm{k}\sigma}
  \varepsilon_{\bm{k}} c_{\bm{k}\sigma}^{\dag} c_{\bf{k}\sigma}
  + \sum_{\bm{k}\sigma m}
  (V_{m} c_{{\bf k}\sigma}^{\dag}f_{m\sigma}+{\rm h.c.})
  + H_{\rm f},
\end{eqnarray}
where $\varepsilon_{\bm{k}}$ is the conduction electron energy dispersion,
$c_{\bm{k}\sigma}$ is the annihilation operator for conduction
electrons with momentum $\bm{k}$ and spin $\sigma$,
$\sigma$=+1 ($-$1) for up (down) spin,
$f_{m\sigma}$ is the annihilation operator for $f$-electrons
with spin $\sigma$ and angular momentum $m$(=$-3$,$\cdots$,3),
and $V_{m}$ is the hybridization between $f$ electrons and the conduction band.
In filled skutterudites, the conduction band is given by $a_{\rm u}$,
constructed from $p$-orbitals of the pnictogens \cite{Harima1}.
Note that hybridization occurs
between states with the same symmetry.
Since the $a_{\rm u}$ conduction band has xyz symmetry,
we set $V_{2}$=$-V_{-2}$=$V$ and zero for the other values of $m$.
Throughout this paper, the energy unit is half of the bandwidth
of the conduction band \cite{Harima2}.

The $f$-electron term $H_{\rm f}$ in Eq.~(\ref{AndersonModel})
is given by
\begin{eqnarray}
  && \! H_{\rm f} \!=\!
  \sum_{m,\sigma,m',\sigma} (B_{m,m'}\delta_{\sigma\sigma'}
  \!+\! \lambda \zeta_{m,\sigma,m',\sigma'})
  f_{m\sigma}^{\dag}f_{m'\sigma'} \nonumber \\
  &+& \! \sum_{m_1 \sim m_4} \! \sum_{\sigma_1,\sigma_2}
  \! I_{m_1,m_2,m_3,m_4}
  f_{m_1\sigma_1}^{\dag} \! f_{m_2\sigma_2}^{\dag}
  \! f_{m_3\sigma_2} \! f_{m_4\sigma_1},
\end{eqnarray}
where $B_{m,m'}$ is the CEF potential for
angular momentum $\ell=3$,
$\delta_{\sigma\sigma'}=(1+\sigma\sigma')/2$,
$\lambda$ is the spin-orbit coupling.
The matrix elements are given by
$\zeta_{m,\sigma,m,\sigma}=m\sigma/2$,
$\zeta_{m+1,\mp 1,m, \pm 1}=\sqrt{12-m(m \pm 1)}/2$,
and zero for the other cases.
The Coulomb integral $I_{m_1, m_2, m_3, m_4}$ is expressed
by the combination of four Racah parameters,
$A$, $B$, $C$, and $D$ \cite{Racah}.
For the $T_{\rm h}$ point group, $B_{m,m'}$ is given by three parameters,
$x$, $y$, and $W$ \cite{Takegahara}, where $x$ and $y$ specify
the CEF scheme and $W$ denotes the energy scale for the CEF potential.

\begin{figure}[t]
\includegraphics[width=1.0\linewidth]{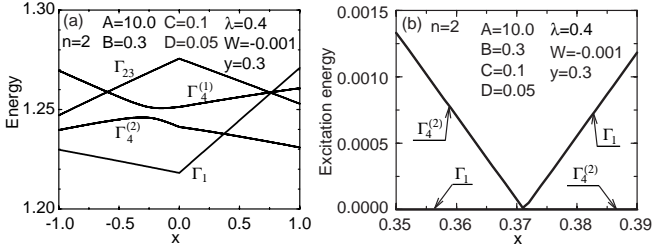}
\caption{(a) Eigen-energies of $H_{\rm f}$ vs. $x$ for $n$=2.
(b) Excitation energy vs. $x$ around at $x$$\sim$0.37.}
\end{figure}

Let us first examine the local $f$-electron states,
in order to set the parameters and check the validity of $H_{\rm f}$.
In Fig.~1(a), eigen-energies of $H_{\rm f}$ in the low-energy region
are plotted as functions of $x$ for $n$=2.
The other CEF parameters, Racah parameters, and the
spin-orbit interaction are appropriately chosen for filled skutterudites.
It is observed that the results for the $LS$ coupling scheme are
well reproduced, but we also emphasize that $H_{\rm f}$ correctly reproduces
the local $f$-electron states for $both$ the $LS$ and $j$-$j$ coupling schemes,
depending on the Coulomb interactions and spin-orbit coupling,
for $any$ value of $n$.
Recall that the $\Gamma_1$ singlet ground state and $\Gamma_4^{(2)}$ triplet
excited state are split by only a small energy difference
in Pr-based filled skutterudites \cite{Gamma1}.
To study this, we magnify the region in which
the $\Gamma_1$ and $\Gamma_4^{(2)}$ states cross.
In the following, we consider the region in Fig.~1(b) where
$0.36 \alt x \alt 0.37$.

Here we include the hybridization between $f$ electrons and conduction
band states. For this purpose, we employ the numerical
renormalization group (NRG) method \cite{NRG},
in which the momentum space of the conduction electrons
is logarithmically discretized near the Fermi energy.
The discretization is determined by a cut-off parameter $\Lambda$.
In this paper we use $\Lambda$=5.
Due to the limitation of computer memory, we keep only $4000$
low-energy states for each renormalization step, but
the results do not change qualitatively for $\Lambda$$\ge$3
when 4000 states are kept.

\begin{figure}[t]
\includegraphics[width=1.0\linewidth]{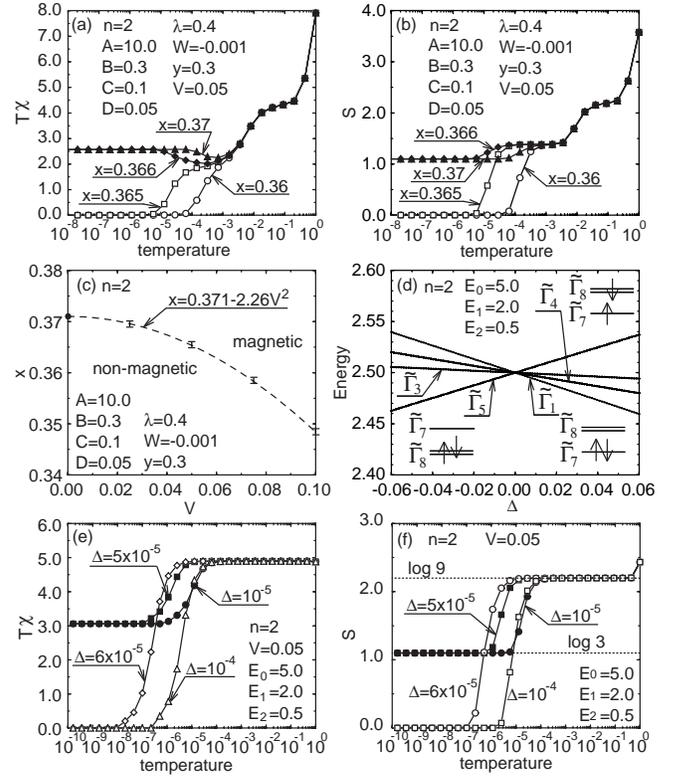}
\caption{(a) Magnetic susceptibility and (b) entropy of $f$ electron
vs. $T$ for $n$=2 in the Anderson model (\ref{AndersonModel})
for $x$=0.36, 0.365, 0.366, and 0.37.
(c) Critical values of $x$ between magnetic and non-magnetic regions
for $V$=0.025, 0.05, 0.075 and 0.1.
Broken curve is drawn to fit the numerical results.
(d) Eigen-energies of ${\tilde H}_{\rm f}$ for $n$=2.
Insets denote $f$-electron configurations for ${\tilde \Gamma}_1$,
${\tilde \Gamma}_4$, and ${\tilde \Gamma}_5$ states, in which
two $f$ electrons are accommodated in the $f^1$-energy levels.
(e) Magnetic susceptibility and (f) entropy of $f$ electron vs. $T$
for $n$=2 in the Anderson model (\ref{AndersonModel2})
for various values of $\Delta$.}
\end{figure}

In order to examine the properties of the $f$ electrons,
we evaluate the magnetic susceptibility $\chi$ and the entropy $S$
of the $f$ electrons.
In Fig.~2(a), we show calculated results for $T\chi$
for $n$=2 and several values of $x$ between 0.36 and 0.37.
For $x \alt 0.365$, $T\chi$ goes to zero for small $T$,
while for $x \agt 0.366$, $T\chi$ becomes constant
at low temperatures \cite{Hotta2}.
As shown in Fig.~2(b) for $S$, we obtain $\log 3$
as residual entropy for $x$=0.366 and 0.37,
indicating that a local triplet remains in the low-temperature region.
Here we emphasize that magnetic fluctuations still remain
at low temperatures, even when $\Gamma_1$ is the ground state,
if $\Gamma_4^{(2)}$ is an excited state with a small excitation energy.
We suggest that anomalous properties of Pr-based filled skutterudites
originate from such magnetic fluctuations.

We remark that the transition between magnetic
and non-magnetic states is governed by the exchange
interaction $J_{\rm cf}$ between $f$ electrons and the conduction band,
expressed as $J_{\rm cf}$=$V^2/\delta \! E$, where $\delta \! E$
denotes the energy difference between the $f^2$ and $f^3$
(or $f^1$) lowest-energy states.
As shown in Fig.~2(c), 
the boundary curve between magnetic and non-magnetic phases
is proportional to $V^2$,
suggesting that the dominant energy scale should be $J_{\rm cf}$
for the appearance of magnetic fluctuations.

Note that the transition between magnetic and non-magnetic
states seems to be abrupt, not gradual as observed in the usual
Kondo system and the two-channel Anderson model \cite{2chan}.
In the Kondo problem, the local moment is suppressed
$only$ by hybridization with conduction electrons,
but in the present case, a singlet ground state is also
obtained through local level crossing due to the CEF potential,
in addition to the hybridization process.
Furthermore, localized orbitals exist in the present model,
as will be discussed later.
Namely, the magnetic moments of the $f$ orbitals hybridized with
the conduction band are suppressed as in the Kondo effect, while
the moments of the localized orbitals are not screened \cite{comment1}.
An abrupt change in the magnetic properties is caused by
the duality of the $f$ orbitals in combination with 
the level crossing effect due to the CEF potential.

In order to understand the above point more explicitly,
it is useful to express the $f$-electron state with
xyz symmetry in the $j$-$j$ coupling scheme as
\begin{equation}
 |{\rm xyz},\sigma \rangle=
 \sqrt{4/7}|7/2,{\tilde \Gamma}_7,{\tilde \sigma} \rangle-
 \sqrt{3/7} |5/2,{\tilde \Gamma}_7,{\tilde \sigma} \rangle,
\end{equation}
where $|j,{\tilde \Gamma},{\tilde \sigma} \rangle$
denotes the state in the $j$-$j$ coupling scheme
with total angular momentum $j$,
irreducible representation ${\tilde \Gamma}$ for the
$O_{\rm h}$ point group \cite{comment2},
and pseudospin ${\tilde \sigma}$ \cite{comment3}.
This relation indicates that
only the ${\tilde \Gamma}_7$ state is hybridized with
$a_{\rm u}$ conduction band states with xyz symmetry,
leading to the Kondo effect,
while the ${\tilde \Gamma}_8$ electrons are localized
and become the source of local fluctuations at low temperatures.
It is an important issue of filled skutterudite structures
that the nature of the $f$ electrons can clearly be distinguished
as itinerant ${\tilde \Gamma}_7$ or localized ${\tilde \Gamma}_8$.
This point provides a possible explanation for the heavy-fermion
phenomenon occurring in $f^2$-electron systems such as
Pr-based filled skutterudites.

In the $j$-$j$ coupling scheme, the model is rewritten as
\begin{eqnarray}
  \label{AndersonModel2}
  {\tilde H} = \sum_{\bm{k} {\tilde \sigma}}
  E_{\bm{k}} a_{\bm{k}{\tilde \sigma}}^{\dag} a_{\bf{k}{\tilde \sigma}}
  +\sum_{\bm{k}{\tilde \sigma}}
  (V a_{\bm{k}{\tilde \sigma}}^{\dag} {\tilde f}_{c{\tilde \sigma}}
  +{\rm h.c.}) + {\tilde H}_{\rm f},
\end{eqnarray}
where $E_{\bm{k}}$ is the dispersion of the ${\tilde \Gamma}_7$
conduction electrons,
$a_{\bm{k}{\tilde \sigma}}$ is the annihilation operator
for conduction electrons
with momentum $\bm{k}$ and pseudospin ${\tilde \sigma}$,
${\tilde f}_{\gamma{\tilde \sigma}}$ is the annihilation operator
for $f$ electrons with pseudospin ${\tilde \sigma}$ and
orbital $\gamma$ for the $j$=5/2 sextet on the impurity site, and
$V$ is the hybridization between conduction and $f$ electrons
with ${\tilde \Gamma}_7$ symmetry.
The orbital index $\gamma$ distinguishes three kinds of
Kramers doublets, two ${\tilde \Gamma}_8$ ($a$ and $b$) and
one ${\tilde \Gamma}_7$ ($c$).
The local $f$-electron term ${\tilde H}_{\rm f}$ is given by
\begin{eqnarray}
{\tilde H}_{\rm f}
  &=& \sum_{\gamma,{\tilde \sigma}} {\tilde B}_{\gamma}
  {\tilde f}_{\gamma{\tilde \sigma}}^{\dag}
  {\tilde f}_{\gamma{\tilde \sigma}}
  +(1/2) \sum_{\gamma_1 \sim \gamma_4}
  \sum_{{\tilde \sigma}_1,{\tilde \sigma}_2}
  {\tilde I}^{{\tilde \sigma}_1,{\tilde \sigma}_2}
  _{\gamma_1, \gamma_2, \gamma_3, \gamma_4} \nonumber \\
  &\times&
  {\tilde f}_{\gamma_1{\tilde \sigma}_1}^{\dag}
  {\tilde f}_{\gamma_2{\tilde \sigma}_2}^{\dag}
  {\tilde f}_{\gamma_3{\tilde \sigma}_2}
  {\tilde f}_{\gamma_4{\tilde \sigma}_1},
\end{eqnarray}
where ${\tilde B}_{\gamma}$ is the CEF potential and
${\tilde I}$ is the Coulomb interaction
in the $j$-$j$ coupling scheme, expressed by other
Racah parameters $E_0$, $E_1$, and $E_2$ \cite{Hotta1}.

Since the CEF potential is already diagonalized,
it is convenient to introduce a level splitting $\Delta$
between ${\tilde \Gamma}_7$ and ${\tilde \Gamma}_8$ as
$\Delta$=${\tilde B}_{{\tilde \Gamma}_8}$$-$
${\tilde B}_{{\tilde \Gamma}_7}$.
In Fig.~2(d), we depict the low-energy states of ${\tilde H}_{\rm f}$
vs. $\Delta$ for $n$=2.
The ground state for $\Delta$$<$0 is a ${\tilde \Gamma}_5$ triplet
composed of a couple of ${\tilde \Gamma}_8$ electrons,
while for $\Delta$$>$0, it is a ${\tilde \Gamma}_1$ singlet
which is mainly composed of two ${\tilde \Gamma}_7$ electrons.
Note that for $\Delta$$>$0, the first excited state is a
${\tilde \Gamma}_4$ triplet composed of ${\tilde \Gamma}_7$
and ${\tilde \Gamma}_8$ electrons.

Now we again employ the NRG method to investigate the magnetic
properties of the Anderson model in the $j$-$j$ coupling scheme.
In Figs.~2(e) and (f), we depict $T\chi$ and $S$
for several values of $\Delta$.
Racah parameters are set as appropriate values for rare-earth
compounds.
It is observed that $T\chi$ becomes constant
and residual entropy $\log 3$ remains
at low temperatures for $\Delta \alt 5 \times 10^{-5}$,
while $T\chi$ goes to zero rapidly
for $\Delta \agt 6\times10^{-5}$.
Note that the transition is also dominated by $J_{\rm cf}$.

Let us consider the reason why the essential magnetic features
of Pr-based filled skutterudites can be captured as described
in the $j$-$j$ coupling scheme.
As shown in Fig.~2(d),
when the ground state is a ${\tilde \Gamma}_1$ singlet,
there are two triplet excited states,
${\tilde \Gamma}_4$ and ${\tilde \Gamma}_5$.
The ${\tilde \Gamma}_5$ triplet is composed of a pair of
${\tilde \Gamma}_8$ electrons, while the ${\tilde \Gamma}_4$
triplet is composed of one
${\tilde \Gamma}_7$ and one ${\tilde \Gamma}_8$ electron.
Since ${\tilde \Gamma}_8$ electrons
do $not$ hybridize with ${\tilde \Gamma}_7$ conduction electrons,
the ${\tilde \Gamma}_5$ triplet can survive.
Note that $\Gamma_4$ triplets in $T_{\rm h}$ are given by
the mixtures of ${\tilde \Gamma}_4$ and ${\tilde \Gamma}_5$
in $O_{\rm h}$.
Such a mixing is not included in the $j$-$j$ coupling scheme,
but there is a ${\tilde \Gamma}_5$ excited state.
Thus, the local triplet still remains, as long as
the excitation energy is smaller than $J_{\rm cf}$.


Thus far, we have focused on the case of $n$=2, but we can
further study the cases of $n$=1$\sim$13
based on the original Anderson model (\ref{AndersonModel}).
In the point-charge picture, the CEF parameters do $not$
depend on $n$, since the CEF effect is due to electrostatic potential
from ligand ions surrounding the rare earth.
In the following, we fix $x$=0.366,
in which the local $f^2$ ground state is a $\Gamma_1$ singlet,
but for which we have found significant magnetic fluctuations at low
temperatures.
In Figs.~3(a) and (b), the magnetic susceptibility and entropy
for each case of $n$$<$7 are shown.
For $n$=2, 4, and 6, the ground state is a $\Gamma_1$ singlet.
Except for the case of $n$=2, the excitation energy is large and
both magnetic and orbital fluctuations should be
rapidly suppressed with decreasing temperature.
Thus, $T\chi$ immediately becomes zero at low temperatures
for $n$=4 and 6.
On the other hand, for $n$=1, 3, and 5, the ground state is a
$\Gamma_{67}$ quartet in $T_{\rm h}$
(=${\tilde \Gamma}_8$ in $O_{\rm h}$),
as confirmed from the residual entropy of $\log 4$.

\begin{figure}[t]
\begin{center}
\includegraphics[width=1.0\linewidth]{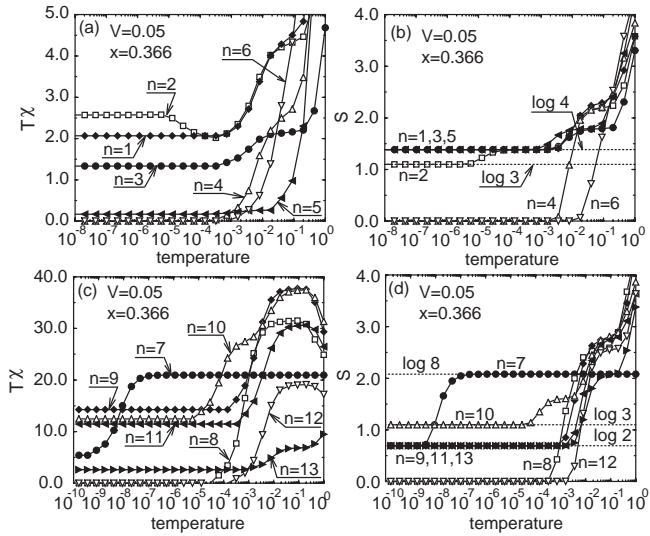}
\caption{(a) Magnetic susceptibility and (b) entropy
of $f$ electrons vs. $T$ for $n$=1$\sim$6.
(c) Magnetic susceptibility and (d) entropy
for $n$=7$\sim$13. We set $x$=0.366 and all other parameters are the
same as those in Figs.~2 (a) and (b).}
\end{center}
\end{figure}

In Figs.~3(c) and (d), we show the NRG results for susceptibility
and entropy for $n$$\ge$7.
First of all, the absolute values of $\chi$ are much larger than
those for $n$$<$7, since the total angular momentum $J$ becomes large
for $n$$\ge$7 owing to Hund's rule coupling.
Typically, at half-filling, total spin $S$(=$J$) is equal to 7/2,
and the Curie constant for an isolated ion is as large as
21$\mu_{\rm B}^2/k_{\rm B}$.
Over a broad temperature region, this value has been observed
in Fig.~3(c) for $n$=7, indicating that the $S$=7/2 spin survives
at relatively low temperatures.
In fact, we clearly see an entropy of $\log 8$ from the $S$=7/2 octet,
as shown in Fig.~3(d) \cite{comment4}.
For the cases of $n$=8 and 12, the ground state is a $\Gamma_1$ singlet
and the magnetic excited state energy is now large,
in sharp contrast with the case of $n$=2.
Thus, the susceptibility rapidly goes to zero.
For $n$=9, 11, and 13, the local ground state is a $\Gamma_5$ doublet
in $T_{\rm h}$ (=${\tilde \Gamma}_6$ in $O_{\rm h}$).
Thus, this state does not hybridize with the conduction band and
the magnetic moment still persists even in the low-temperature region.
In fact, we observe a residual entropy of $\log 2$ in these cases,
as shown in Fig.~3(d).
For $n$=10, the local ground state is a $\Gamma_4$ triplet,
but as observed in Figs.~3(c) and (d),
the local triplet seems to remain at low temperatures.
This is easily understood, if we recall that 
$\Gamma_4$ triplets of $T_{\rm h}$ are given by
mixtures of ${\tilde \Gamma}_4$ and ${\tilde \Gamma}_5$
of $O_{\rm h}$.
As mentioned above, the ${\tilde \Gamma}_5$ triplet still persists
even after hybridization, 
since a ${\tilde \Gamma}_8$ electron
does $not$ hybridize with a ${\tilde \Gamma}_7$ conduction electron.


In summary, we have analyzed a multi-orbital Anderson model
using the NRG method.
We have established that ${\tilde \Gamma}_7$ states hybridize with
the conduction band, while ${\tilde \Gamma}_8$ states are localized.
For $n$=2, corresponding to Pr-based filled skutterudites,
it has been found that magnetic fluctuations significantly persist,
if there is a $\Gamma_4^{(2)}$ excited state triplet
with a small excitation energy.
It has been shown that essentially these same results are
reproduced by the Anderson model in the $j$-$j$ coupling scheme.
We believe that these results open a new door for the study of magnetic
and superconducting properties of Pr-based filled skutterudites
from a microscopic point of view.


The author thanks H. Harima, K. Kubo, H. Onishi, and K. Takegahara
for discussion.
He also thanks R. E. Walstedt for critical reading of the manuscript.
He is supported by Grant-in-Aids for Scientific
Research of Japan Society for the Promotion of Science
and for Scientific Research in Priority Area ``Skutterudites''
of the Ministry of Education, Culture, Sports,
Science, and Technology of Japan.
The computation has been done using the facility
of the Supercomputer Center of Institute for Solid State Physics,
University of Tokyo.



\begin{thebibliography}{99}
\bibitem{comment1}
In actuality, due to conduction bands other than $a_{\rm u}$
which hybridize with ${\tilde \Gamma}_8$ states,
the transition may be gradual, but the anisotropic
hybridization is still important.
\bibitem{comment2}
In the $j$-$j$ coupling scheme, we use the notation for $O_{\rm h}$,
since the CEF potential for one $f$-electron state in the $j$=5/2
sextet is the same for $O_{\rm h}$ and $T_{\rm h}$.
See also Ref.~\cite{Takegahara}.
\bibitem{comment3}
There is one-to-one correspondence
between $\sigma$ and ${\tilde \sigma}$.
\bibitem{comment4}
At extremely low temperatures, however, $S$=7/2 spin seems to be
eventually screened and only $S$=1/2 remains,
leading to the residual entropy $\log 2$.
\end{thebibliography}
\end{document}